\documentclass[floatfix,aps,twocolumn,prb,amsmath,amssymb,showpacs,superscriptaddress]{revtex4-1}

\usepackage{eurosym}
\usepackage{amssymb}
\usepackage{amsmath}
\usepackage{amsfonts}
\usepackage{amsbsy}
\usepackage{graphicx}
\usepackage{bm}
\usepackage{color}
\usepackage{hyperref}
\usepackage{units}

\begin{document}

\title{Topological quantum phase transition in individual Fe Atoms on MoS$_2$/Au(111)}

\author{G. G. Blesio}
\affiliation{Jo\v{z}ef Stefan Institute, Jamova 39, SI-1000 Ljubljana, Slovenia}

\author{A. A. Aligia}
\affiliation{Instituto de Nanociencia y Nanotecnolog\'{\i}a CNEA-CONICET,
Centro At\'{o}mico Bariloche and Instituto Balseiro, 8400 Bariloche, Argentina}

\begin{abstract}

In a recent experiment [Trishin {\it et al.}, Phys. Rev. Lett. \textbf{127}, 236801 (2021)] a rich physics was observed for Fe atoms on 
MoS$_2$/Au(111), characterized by three different behaviors depending 
on the spectral density of the substrate $\rho_c$: one dominated by a single ion anisotropy, one by the Kondo effect, and an intermediate one. 
Based on symmetry and previous works, we show that the 
appropriate model to describe the system is the anisotropic two-channel spin-1  Kondo model (A2CS1KM), which has an underlying
topological quantum phase transition (TQPT) between an ordinary Fermi liquid and a topological one with a non-trivial value of
the Luttinger integral. 
Solving the model with the numerical renormalization group (NRG), we show that the different behaviors can be explained in a unified fashion as a function of $\rho_c$, and correspond to both Fermi liquids and an intermediate
regime close to the TQPT in the topological phase.
Further experiments should confirm this transition. 
\end{abstract}


\maketitle

The understanding of the basic physics of systems composed of
magnetic atoms or molecules on metallic surfaces is a subject 
of great interest in the past two decades.
This is motivated by possible promising technological applications in the field of spintronics \cite{wolf}, for which
coherent manipulation of the spin is essential \cite{yang1}, and 
also as a test for paradigmatic theories in condensed matter,
such as the Kondo effect \cite{hewson97,kondo64}. In its simplest form, the Kondo effect arises when the free electrons of a metallic host
screen completely the magnetic moment of an impurity \cite{hewson97}.

Several relevant experiments using a scanning-tunneling microscope 
(STM) have studied the low-energy electronic structure of systems in which magnetic units were deposited on noble-metal surfaces. 
The magnetic units were either transition-metal atoms with 
incomplete $d$ shells \cite{coau,mano,naga,knorr,wahl04,limot,neel,vita,choiPRL,choi}, 
or molecules containing transition-metal atoms (or other localized electrons) 
\cite{wahl,gao07,roch,parks,mina,zhang,kugel14,karan,esat,iancu,hira,ormaza17,orma3,verlhac,yang,guo,meng,gao}.
In most of these works, several variants of the Kondo effect are apparent. The theoretical treatments become involved because of the 
need to treat accurately the Coulomb repulsion in the incomplete $d$ shells and in many cases, several localized electrons and 
conduction bands are relevant at low energies, leading to non-trivial
interference effects that modify the line-shape in the 
differential conductance $G(V)=dI/dV$, where $I$ is the current and 
$V$ the voltage applied to the STM tip 
\cite{uj,meri,mirages,lin,trimer,serge,frank,morr,interf,lim}.

One of the simplest models when more that one $d$ orbital is involved, is the Anderson model with just two localized orbitals coupled by Hund rules to form a spin 1, hybridized with 
conduction states of the same symmetry, including anisotropy. In the integer valence limit, the model reduces to the 
A2CS1KM 
with hard axis anisotropy $D$. Rather surprisingly, 
it was found a few years ago that these models 
have a topological quantum phase transition (TQPT) between an ordinary Fermi liquid (FL) when $D$ is below a critical value $D_c$ to a topological Fermi liquid (TFL) for $D>D_c$ \cite{blesio18,blesio19}. At the TQPT,
the Luttinger integral jumps from 0 (trivial)  to $\pi/2$ (topological) with increasing $D$ and therefore, for $D>D_c$, the system cannot be adiabatically connected to 
a non-interacting system, breaking Landau's hypothesis. 
For this reason the TFL has previously been called ``non-Landau'' FL.

\begin{figure}[hb]
\begin{center}
\includegraphics*[width=\columnwidth]{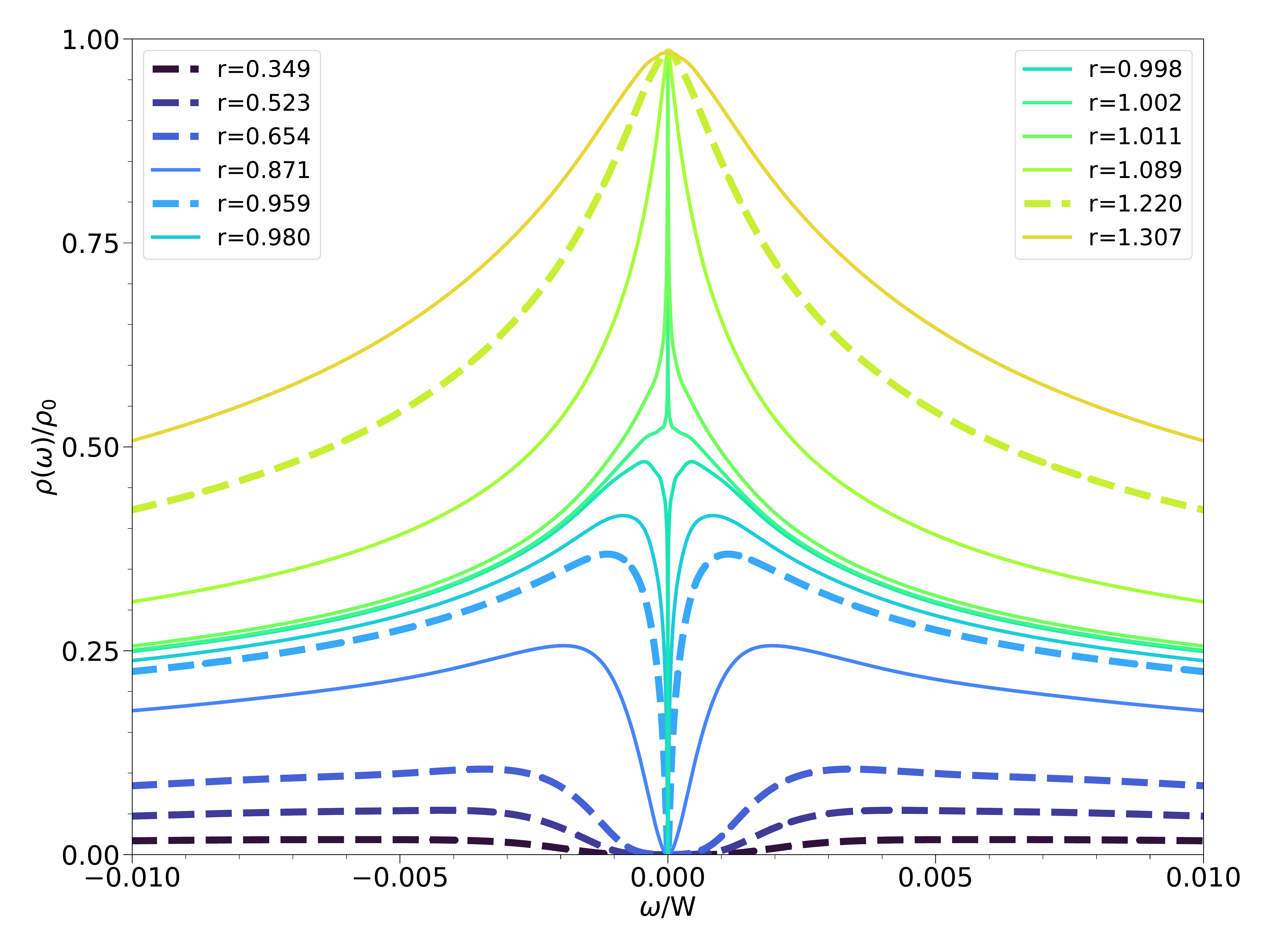}
\end{center}
\caption{(Color online) Spectral density 
of localized electrons 
of the A2CS1KM as a function of frequency for several values 
of $r=J^{\prime}/J^{\prime}_c$, where $J^{\prime}_c$ 
is the value of $J^{\prime}$ at the TQPT.
$\rho_0$ is the value of $\rho(0)$ 
predicted by the ordinary Friedel sum rule
with vanishing Luttinger integral \cite{nl}. 
Thicker dashed lines correspond to the values of $r$ 
chosen for the different curves in Fig. 2.}
\label{fig1}
\end{figure}

For equivalent channels $D_c \sim 2.5 T_K^0$, where $T_K^0$ 
is the Kondo temperature for $D=0$, which increases 
with increasing $J^{\prime}=\rho_c J$, where $\rho_c$
is the density of conduction states and $J$ the exchange coupling.
The evolution of the spectral density for localized states 
$\rho(\omega)$ with $D$ for constant  
$T_K^0$ is shown in Fig.  11 of Ref. \onlinecite{blesio19} 
and in Fig. \ref{fig1} below for constant $D$ and changing 
$J^{\prime}$.
For small $D/T_K^0$, the spectral density has the usual Kondo peak.
Increasing $D/T_K^0$, near the TQPT, $\rho(\omega)$ displays
a very narrow peak mounted on a broader peak (the latter characteristic of the two-channel spin-1/2 Kondo model \cite{blesio18}). 
At the TQPT, the narrow peak changes suddenly to a dip as a consequence of the jump of the 
Luttinger integral. 
For large $D/T_K^0$, the spectral density has two steps at 
$\omega=\pm D$, characteristic of inelastic transitions
(the broadening in Fig. \ref{fig1} for large $\omega$ is an 
artifact of the NRG).
Similar behavior has been
found in other models \cite{leo,curtin18,nishi}.

This theory has been shown to explain in a consistent way 
relevant experiments that have been interpreted previously 
using alternative (and some times implausible) explanations.
For the system of an isolated iron phthalocyanine
molecule deposited on the Au(111) surface [FePc/Au(111)] a 
narrow dip in $G(V)$ near $V=0$ was observed \cite{mina}, 
mounted on a broader peak, which broadens as the Kondo interaction is weakened  \cite{hira} and is transformed into a narrow peak 
with an applied magnetic field \cite{yang}. These three 
experiments were explained using the A2CS1KM 
with two inequivalent channels \cite{nl}.
A detailed justification of this model for FePc/Au(111) is given
in Ref. \onlinecite{sb}. MnPc/Au(111) also shows the transformation
of the narrow dip to a narrow peak as a function of an applied magnetic field \cite{guo}. The authors proposed an interpretation in terms of a quantum phase transition involving localized singlet states \cite{guo}. 
However, this requires that the singlet be below the triplet by a few
meV, while in fact the triplet is energetically favored by a Hund's coupling of the order of 1 eV \cite{mina}. 

In the system of nickelocene on Cu(100) [Nc/Cu(100)], as the STM tip 
is approached to the molecule, increasing hybridization between 
localized and conduction electrons (effectively increasing $T_K^0$),
a transition from a dip in $G(V)$ to a peak is observed \cite{ormaza17,mohr} which was discontinuous in 2/3 of the cases. 
The authors have tentatively ascribed the change of behavior to a crossover in the spin of the molecule from 1 in the tunneling 
regime to 1/2 in the contact regime. However, the neglected dynamical correlations should lead to a singlet ground state in both cases,  and the change in the molecular charge is actually insufficient to account for the large change in the magnetization. Very recently,
the different behaviors were explained in terms of the above mentioned TQPT \cite{nc}. Similar experiments were carried out for iron porphyrin molecules on Au(111) \cite{meng,gao}. In Ref. \onlinecite{meng} the dip narrows in the contact regime instead of turning to a peak, what points out to a weaker hybridization of the localized states with the substrate in comparison to Nc/Cu(100) \cite{meng}. However the transformation from a dip to a peak 
is observed using a Br decorated Au(111) surface \cite{gao}.

A variant of the interaction between magnetic units and 
metallic surfaces has been realized introducing thin decoupling
layers between both, weakening the hybridization between 
localized and conduction states \cite{hein,hir,loth,paul,tri}.
In particular, Trishin \textit{et al.} considered a system 
consisting of an Fe atom on top of a monolayer of MoS$_2$ deposited
in turn on a Au(111) surface \cite{tri}. Particularly 
interesting in this system is the fact that MoS$_2$ on Au(111)
forms a Moir\'e structure, which implies strong local variations
of the density of conduction electrons. 
Therefore, depending on the 
specific position at which the Fe adatom is located, dramatic
variations of the effective exchange between electron localized
at the Fe adatom and conduction electrons in the rest 
of the system takes place. The differential conductance $G(V)$ 
were recorded on nearly 40 different Fe atoms and six of them
are presented in Ref. \onlinecite{tri}. 
We denote the corresponding spectra by the letters (a) to (f)
used in Fig. 2  of Ref. \onlinecite{tri}.
Five of them [(b) to (f)] are reproduced here in Fig. 2. 
The spectra  (a) to (d) contain a dip of
variable width in $G(V)$ near $V=0$. Spectrum (e) contains 
a narrow dip mounted on a broader peak, and spectrum (f)
consists of a single Kondo peak.

Taking into account that the rigorous study of the A2CS1KM is
both recent and difficult, it is understandable that the 
theoretical part of Ref. \onlinecite{tri} was based on more
traditional one-channel approaches, to explain the different regimes 
[perturbation theory for (b) to (d) \cite{ternes}, Lorentz peak plus Frota dip (without justification) for (e) and Frota peak for (f)]. 
However, not only more than one channel is expected
(see below) but also a recent experimental study of the same 
group for Mn atoms,
concludes that more than one channel is coupled to the 
atom \cite{tri2}. Since Mn is next to Fe in the periodic table,
one expects the same to be valid for Fe atoms.

In this work we show that all spectra can be semiquantitatively
explained using the A2CS1KM with equivalent channels, fixed 
$D$ and varying $J^{\prime}=\rho_c J$, 
assumed the same for both channels. The 
minor discrepancies are likely due to the dependence
on the energy of the conduction density of states $\rho_c$ 
which we neglect.
Since alternative perturbative or diagrammatic approximations
fail to describe the TQPT \cite{blesio19}, we use the NRG.

The model for the system can be justified on general physical grounds.
Density-functional-theory (DFT) calculations of Fe atoms on free-
standing MoS$_2$ indicate that the spin state of the atoms
is either $S=1$ \cite{wang} or $S=2$ \cite{chen}. 
However, for $S=2$, one would 
expect a second jump in $G(V)$ at larger $|V|$ in the regime of 
low $J$, which is not observed experimentally 
[Fig. 3(h) of Ref. \onlinecite{tri}]. 
In addition, experiment as 
well as the DFT calculations, indicate that the Fe atom are located
in positions with symmetry corresponding to the point group $C_{3v}$.
Therefore, the Fe $3d$ orbitals are split in one $A_{1}$ singlet 
and two $E$ doublets \cite{tkro}. Choosing the coordinates in such a way that $z$ is perpendicular to the surface, the $3d$
orbital with symmetry $3z^{2}-r^{2}$ transforms as $A_{1}$, 
one $E$ doublet has the form \cite{tkro}

\begin{eqnarray}
|x\rangle  &=&\alpha |xz\rangle +\beta |\left( x^{2}-y^{2}\right) /2\rangle ,
\notag \\
|y\rangle  &=&\alpha |yz\rangle -\beta |xy\rangle,  
\label{e}
\end{eqnarray}
and the other $E$ doublet has the same form interchanging $\alpha$ with $-\beta$ (so that all states are orthogonal).

It is likely that the  spin 1 is formed occupying the two states 
of an $E$ doublet. Then it is expected that the spin-orbit coupling
(SOC) originates a hard axis anisotropy $D(S_z)^2$ with $D>0$ \cite{nc} and that each of the degenerate orbitals of the doublet 
hybridizes with conduction states of the same symmetry \cite{blesio18,blesio19,nc}. In the integer valence limit, 
this reasoning leads naturally to the A2CS1KM with the same 
exchange interaction for both channels. 
In the other case in which one of the orbitals of the 
spin 1 corresponds to the $A_1$ symmetry, the SOC favors a particular 
linear combination of the $E$ states for the other orbital \cite{sb}
and the low-energy model is also the A2CS1KM but with different 
exchange interactions for both symmetries. This case corresponds
to iron phthalocyanine on Au(111) \cite{nl,sb}.  For the system
studied here, we find that very different exchange constants
cannot explain the experiments, and then we assume that
they are equal. Therefore, the model takes the form

\begin{eqnarray}
H_{K} &=&\sum_{k\tau \sigma }\varepsilon _{k\tau }c_{k\tau \sigma }^{\dagger
}c_{k\tau \sigma }+\sum_{k\tau \sigma \sigma ^{\prime }}\frac{J}{2}
c_{k\tau \sigma }^{\dagger }(\vec{\sigma})_{\sigma \sigma ^{\prime
}}c_{k\tau ^{\prime }\sigma ^{\prime }}\cdot \vec{S}  \notag \\
&&+DS_{z}^{2},  \label{hk}
\end{eqnarray}
where $c_{k\tau \sigma }^{\dagger }$ creates a conduction electron 
with point-group symmetry $\tau $, spin $\sigma $ and remaining 
quantum numbers $k$.
The first term describes the substrate bands, the second the Kondo exchange with the localized spin $\vec{S}$ with exchange
couplings $J$, and the last term is the single-ion
uniaxial magnetic anisotropy.

The numerical calculations were performed with the Ljubljana code of the NRG \cite{zitko09,nrglj}. We assume  flat conduction bands extending from $-W$ to $W$ for both symmetries. The density of these states is therefore 
$\rho_c=1/(2W)$. We take $W=1$ eV, $D=2.7$ meV and assume that $J$ varies
depending on the Fe position. In the real system $\rho_c$ depends on 
the position of the Fe adatom, but the relevant parameter is the adimensional 
product $J^{\prime}=\rho_c J$.
The TQPT is at $J^{\prime}_c \sim 0.135$.  
For the sake of clarity, the different curves are shown
for different values of the ratio $r=J^{\prime}/J^{\prime}_c$, which is the main variable in what follows.

We assume that the structure at low voltage $V$ in the 
differential conductance $G(V)=dI/dV$, is determined by the localized
and conduction electrons of symmetry $\tau$ included in the model, and that the STM tip senses mainly the localized $3d$ states with some admixture of
conduction states. Then, the contribution of the model to $G(V)$ 
at zero temperature is (except for a factor $f$ included below) \cite{nl}

\begin{eqnarray}
G_m(V)= 
-\left[ (1-q^{2})\text{Im}G_{\tau \sigma }^{d}(\omega )
+2q\text{Re}G_{\tau \sigma }^{d}(\omega )\right] ,
\label{gm}
\end{eqnarray}
where the $G_{\tau \sigma }^{d}(\omega )$ is the Green function of localized 
electrons for symmetry $\tau$ and spin $\sigma$ (which depends only
on $J^{\prime}$)  
and $q$ is a measure of the contribution of the conduction states.
In the experimental spectrum, there is also a linear background
due to the contribution of other states, and $G_m(V)$ is affected by a 
factor $f$ which depends on the distance of the STM tip to the system.
Thus we take 

\begin{equation}
G(V)=f G_m(V)+A+B\omega,  
\label{g}
\end{equation}
This expression contains only five parameters, but the {\em shape} of each curve
depends mainly on $r$ and to a lesser extent on $q$.
The fact that both $r$ and $q$ increase in going from
(b) to (f) is consistent with the reported greater density 
of conduction states.

\begin{figure}[tbp]
\begin{center}
\includegraphics*[width=0.6\columnwidth]{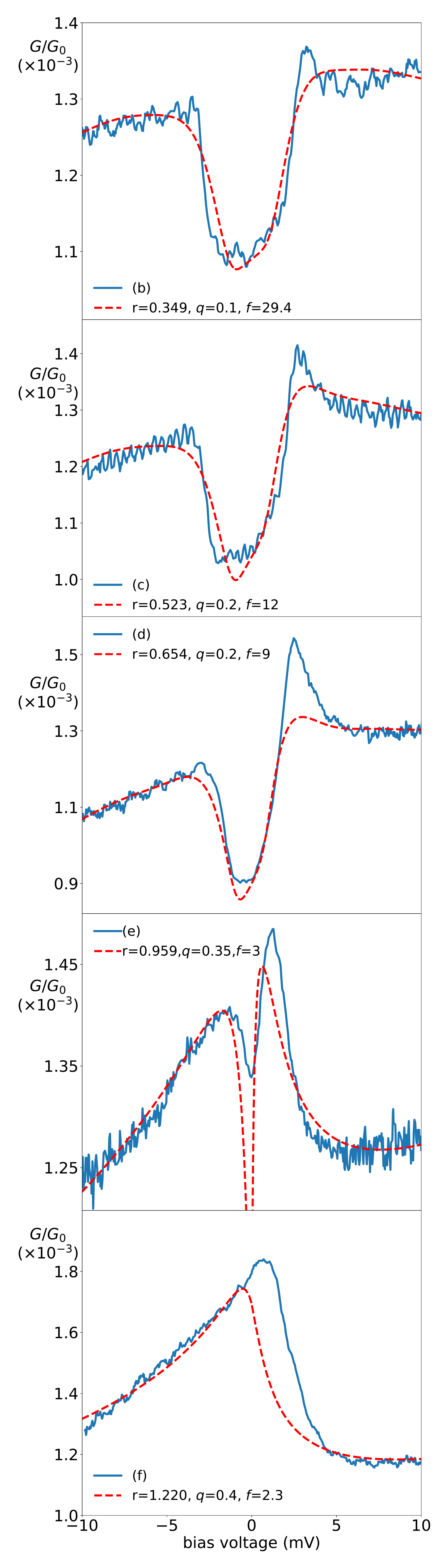}
\end{center}
\caption{(Color online) Differential conductance corresponding to the experimental curves (b) to (f) in Ref. \onlinecite{tri} (blue full lines) and the results for our model (red dashed lines).
Note that (a) is not included and the top figure corresponds to (b).}
\label{fig2}
\end{figure}

In Fig. 2 we represent the experimental curves (b) to (f) obtained from Fig. 2 of Ref. \onlinecite{tri} and the 
corresponding result from our theory. 
We have not included the curve (a), which is similar to a rectangular dip
formed by two step-like functions, because these steps are overbroadened
in our calculations due to technical reasons that limit the resolution of the NRG at large energies \cite{bulla}.
We have not attempted to fit the data
but rather we have chosen  parameters (in particular $r$) that 
provide a semiquantitative agreement. The assumption of a constant 
density 
of conduction states and the effect of other orbitals not contained in our model
should affect a realistic comparison with experiment. Our aim is to show
that the same model can explain the different behaviors observed. 

Experimentally, the figures can be divided into those belonging to a regime dominated by the anisotropy $D$ in which Kondo interactions can be treated pertubatively
[curves (a) to (c)], curve (f) where Kondo exchange $J$ dominates that can be fitted by a simple Frota line shape and an intermediate regime [curves (d) and (e)]. 
For curve (f) a Frota fitting is better than our result, mainly because 
the former has the freedom of locating the position of the peak, which our
theory does not have. This can be improved allowing intermediate valence. 
It is known that a smaller occupancy of the localized states shifts the 
Kondo peak to higher energies. A higher $J^{\prime}$ implies a 
higher hybridization between localized and conduction states and 
this implies a higher degree of intermediate valence. 
However, we want to keep the model simple and our goal is to explain 
all regimes within the same footing rather than to improve minor details for each curve.
The dips obtained by our theory for the curves (d) and (e) are deeper than
the ones experimentally observed. This might also be an effect of intermediate
valence which increases the minimum possible value of the 
localized spectral density \cite{nc}.

The intermediate regime, in particular curve (e) with a narrow dip 
mounted on a broader dip, results naturally from the A2CS1KM 
(see Fig. \ref{fig1} for $r$ slightly smaller than 1),
but is very hard to explain with alternative theories. Before the study of
the A2CS1KM, Minamitani {\it et al.} found a similar behavior in 
$G(V)$ in FePc/Au(111), which was interpreted as a two-stage Kondo effect:
the broad peak would be originated by the Kondo effect for 
one channel ($3z^{2}-r^{2}$ symmetry),
and the narrow dip as the second-stage Kondo effect for states of a different 
symmetry (denoted by $\pi$) \cite{mina,joaq}. A peak indicates dominance
of localized states in the tunneling to the STM tip [low $q$ in Eq. (\ref{g})]
for $3z^{2}-r^{2}$ symmetry and dominance of the tunneling to conduction 
states [large $q$ in Eq. (\ref{g})] for $\pi$ orbitals. 
However with this explanation one would expect a narrowing of both 
features as the molecule is raised from the surface, contrary to more
recent experiments \cite{hira}, which together with the dependence on the 
magnetic field \cite{yang} could be explained using the A2CS1KM \cite{nl}.
For Fe atoms on MoS$_2$/Au(111), the presence of a decoupling layer weakens the tunneling of conduction electrons to the STM tip, and therefore one expects low $q$ (as used in this work) and that
$G(V)$ reflects mostly the density of Fe states. 

To support further the relevance 
of the A2CS1KM to the system, it would be interesting to perform 
experiments for $\rho_c$ slightly larger than
the corresponding value for curve (e) and still smaller than the
corresponding one for
curve (f). In this interval there might be a spectrum 
with a narrow {\em peak} mounted on a broader peak corresponding to 
$r$ slightly larger than $1$. 
In any case, for the curve (e),
an applied magnetic field parallel to the anisotropy direction should turn the dip into a peak in a continuous way \cite{nl}. This is 
another smoking gun for
the A2CS1KM or its extension to intermediate valence, 
since alternative meaningful explanations do not exist.

In summary, we provide strong arguments that indicate that the
appropriate model to describe individual Fe atoms on 
MoS$_2$/Au(111) is the anisotropic two-channel spin-1 
Kondo model, possibly with some degree of intermediate valence in the regime in which the Kondo exchange dominates. 
The different observed curves of differential conductance 
can be explained by the theory modifying the spectral density of the substrate.
The model has a topological quantum phase transition (TQPT)  for anisotropy $D_c \sim 2.5 T_K^0$, where $T_K^0$ is the Kondo temperature for $D=0$. 
We expect that this work stimulates further experiments for conditions near the transition. A fine tuning of the Fe positions 
or changes in the substrate, might
render possible to observe a sudden change from a narrow dip to a narrow peak mounted on a broader peak signaling the TQPT. 
This sudden change has not been observed yet in similar materials.
In any case, applying a magnetic field in the topological phase near the TQPT [curve (e)] should change the narrow dip into a narrow peak confirming the validity of the model.

We thank Felix von Oppen and Katharina Franke for useful discussions. AAA is supported by PICT 2017-2726 and PICT 2018-01546 of the ANPCyT, Argentina. GGB is supported by Slovenian Research Agency (ARRS) under Grant no. P1-0044 and J1-2458. Part of computation were performed on the supercomputer Vega at the Institute of Information Science (IZUM) in Maribor, Slovenia.


\begin{thebibliography}{99}

\bibitem{wolf} S. A. Wolf, D. D. Awschalom, R. A. Buhrman, J. M.
Daughton, S. von Moln\'ar, M. L. Roukes, A. Y. Chtchelkanova,
and D. M. Treger, 
Spintronics: a spin-based electronics vision for the future,
Science \textbf{294}, 1488 (2001).

\bibitem{yang1} K. Yang, W. Paul, S.-H. Phark, P. Willke, Y. Bae, 
T. Choi, T. Esat, A. Ardavan, A. J. Heinrich, and C. P. Lutz, 
Coherent spin manipulation of individual atoms on a surface,
Science \textbf{366}, 509 (2019).

\bibitem{hewson97} A. C. Hewson, \textit{The Kondo Problem to Heavy Fermions}
(Cambridge University Press, Cambridge, UK, 1997). 

\bibitem{kondo64} J. Kondo, 
Resistance Minimum in Dilute Magnetic Alloys, 
Prog. Theor. Phys. \textbf{32}, 37 (1964).

\bibitem{coau} V. Madhavan, W. Chen, T. Jamneala, M. F. Crommie, and N. S. Wingreen,
Tunneling into a Single Magnetic Atom:Spectroscopic Evidence of the Kondo Resonance,
Science \textbf{280}, 567 (1998)

\bibitem{mano} H. C. Manoharan, C. P. Lutz, and D. M. Eigler, 
Quantum mirages formed by coherent projection of electronic structure,
Nature (London) \textbf{403}, 512 (2000).

\bibitem{naga} K. Nagaoka, T. Jamneala, M. Grobis, and M. F. Crommie,
Temperature Dependence of a Single Kondo Impurity,
Phys. Rev. Lett. \textbf{88}, 077205 (2002).

\bibitem{knorr} N. Knorr, M. A. Schneider, L. Diekh\"oner, P. Wahl, and K. Kern, 
Kondo Effect of Single Co Adatoms on Cu Surfaces,
Phys. Rev. Lett. \textbf{88}, 096804 (2002).

\bibitem{wahl04} P. Wahl, L. Diekh\"oner, M. A. Schneider, L. Vitali, G. Wittich, and K. Kern, 
Kondo Temperature of Magnetic Impurities at Surfaces,
Phys. Rev. Lett. \textbf{93}, 176603 (2004).

\bibitem{limot} L. Limot, E. Pehlke, J. Kr\"{o}ger, and R. Berndt,  
Surface-State Localization at Adatoms,
Phys. Rev. Lett. \textbf{94}, 036805 (2005).

\bibitem{neel} N. N\'eel, J. Kr\"oger, L. Limot, K. Palotas, W. A. Hofer, and R. Berndt, 
Conductance and Kondo Effect in a Controlled Single-Atom Contact,
Phys. Rev. Lett. \textbf{98}, 016801 (2007).

\bibitem{vita} L. Vitali, R. Ohmann, S. Stepanow, P. Gambardella, K. Tao, R. Huang, V. S. Stepanyuk, P. Bruno, and K. Kern, 
Kondo Effect in Single Atom Contacts: The Importance of the Atomic Geometry,
Phys. Rev. Lett. \textbf{101}, 216802 (2008).

\bibitem{choiPRL} D.-J. Choi, M. V. Rastei, P. Simon, and L. Limot,
Conductance-Driven Change of the Kondo Effect in a Single Cobalt Atom,  
Phys. Rev. Lett. \textbf{108}, 266803 (2012).

\bibitem{choi} D.-J. Choi, S. Guissart, M. Ormaza, N. Bachellier, O. Bengone, P. Simon, and L. Limot, 
Kondo Resonance of a Co Atom Exchange Coupled to a Ferromagnetic Tip, 
Nano Lett. \textbf{16}, 6298 (2016).



\bibitem{wahl} P. Wahl, L. Diekh\"oner, G. Wittich, L. Vitali, M. A. Schneider, and K. Kern,
Kondo Effect of Molecular Complexes at Surfaces: Ligand Control of the Local Spin Coupling,
Phys. Rev. Lett. \textbf{95}, 166601 (2005).

\bibitem{gao07} L. Gao, W. Ji, Y. B. Hu, Z. H. Cheng, Z. T. Deng, Q. Liu, N. Jiang, X. Lin, W. Guo, S. X. Du, W. A. Hofer, X. C. Xie, and H.-J. Gao, 
Site-Specific Kondo Effect at Ambient Temperatures in Iron-Based Molecules, 
Phys. Rev. Lett. \textbf{99}, 106402 (2007).

\bibitem{roch} N. Roch, S. Florens, V. Bouchiat, W. Wernsdorfer, and F. Balestro, 
Quantum phase transition in a single-molecule quantum dot,
Nature \textbf{453}, 633 (2008).

\bibitem{parks} J. J. Parks, A. R. Champagne, T. A. Costi, W. W. Shum, A. N. Pasupathy, E. Neuscamman, S. Flores-Torres, P. S. Cornaglia, A. A. Aligia, C. A. Balseiro, G. K.-L. Chan, H. D. Abru\~{n}a, and D. C. Ralph,
Mechanical Control of Spin States in Spin-1 Molecules and the
Underscreened Kondo Effect,
Science \textbf{328}, 1370 (2010)

\bibitem{mina} E. Minamitani, N. Tsukahara, D. Matsunaka, Y. Kim, N. Takagi, and M. Kawai, 
Symmetry-Driven Novel Kondo Effect in a Molecule
Phys. Rev. Lett. \textbf{109}, 086602 (2012).

\bibitem{zhang} Y. Zhang, S. Kahle, T. Herden, Ch. Stroh, M. Mayor, U. Schlickum, M. Ternes, 
P. Wahl, and K. Kern, 
Temperature and magnetic field dependence of a Kondo system in the weak coupling regime. 
Nat. Commun. \textbf{4}, 2110 (2013).

\bibitem{kugel14} J. Kügel, M. Karolak, J. Senkpiel, P-J. Hsu, G. Sangiovanni, and M. Bode, 
Relevance of Hybridization and Filling of 3d Orbitals for the Kondo Effect in Transition Metal Phthalocyanines, 
Nano Lett. \textbf{14}, 3895 (2014).

\bibitem{karan} S. Karan, D. Jacob, M. Karolak, C. Hamann, Y. Wang, A.
Weismann, A. I. Lichtenstein, and R. Berndt,
Shifting the Voltage Drop in Electron Transport Through a Single Molecule,
Phys. Rev. Lett. \textbf{115}, 016802 (2015).

\bibitem{esat} T. Esat, B. Lechtenberg, T. Deilmann, C. Wagner, P. Kr\"u ger, R. Temirov, 
M. Rohlfing, F. B. Anders, and F. S. Tautz,
A chemically driven quantum phase transition in a two-molecule Kondo system,
Nat. Phys. \textbf{12}, 867 (2016). 

\bibitem{iancu} V. Iancu, K. Schouteden, Z. Li, and C. Van Haesendonck,
Electron–phonon coupling in engineered magnetic molecules,
Chem. Commun. \textbf{52}, 11359 (2016).

\bibitem{hira} R. Hiraoka, E. Minamitani, R. Arafune, N. Tsukahara, S. Watanabe, M. Kawai, and N. Takagi,
Single-molecule quantum dot as a Kondo simulator,
Nature Commun. \textbf{8}, 16012 (2017).

\bibitem{ormaza17} M. Ormaza, P. Abufager, B. Verlhac, N. Bachellier, M.-L.
Bocquet, N. Lorente, and L. Limot, 
Controlled spin switching in a
metallocene molecular junction, 
Nat. Commun. \textbf{8}, 1974 (2017).

\bibitem{orma3} M. Ormaza, N. Bachellier, M. N. Faraggi, 
B. Verlhac, P. Abufager, P. Ohresser, L. Joly, M. Romeo, F. Scheurer, M.-L. Bocquet, N. Lorente, and L. Limot, 
Efficient spin-flip excitation of a nickelocene
molecule, 
Nano Lett. \textbf{17}, 1877 (2017).

\bibitem{verlhac} B. Verlhac, N. Bachellier, L. Garnier, M. Ormaza, P. Abufager, R. Robles, M.-L. Bocquet, M. Ternes, N. Lorente, 
and L. Limot,
Atomic-scale spin sensing with a single molecule at
the apex of a scanning tunneling microscope,
Science \textbf{366}, 623 (2019).

\bibitem{yang} K. Yang, H. Chen, Th. Pope, Y. Hu, L. Liu, D. Wang, L. Tao, W. Xiao, 
X. Fei, Y-Y. Zhang, H-G Luo, S. Du, T. Xiang, W. A. Hofer, and H-J. Gao, 
Tunable giant magnetoresistance in a single-molecule junction, 
Nature Commun. \textbf{10}, 1038 (2019).

\bibitem{guo} X. Guo, Q. Zhu, L. Zhou, W. Yu, W. Lu, and W. Lian,
Gate tuning and universality of Two-stage Kondo effect in single molecule transistors,
Nature Commun. \textbf{12}, 1566 (2021).

\bibitem{meng} 
X. Meng, J. M\"oller, M. Mansouri, D. S\'anchez-Portal, 
A. Garcia-Lekue, A. Weismann, C. Li, R. Herges, and R, Berndt,
Controlling the Spin States of FeTBrPP on Au(111),
ACS Nano 
https://doi.org/10.1021/acsnano.2c09310

\bibitem{gao} Y. Gao, S, Vlaic, T. Gorni, L. de' Medici, S. Clair, D. Roditchev, and S. Pons,
Manipulation of the magnetic state of a porphyrin-based molecule on gold: From Kondo to quantum nanomagnet via the charge fluctuation regime,
arXiv:2301.01101


\bibitem{uj} O. \'Ujs\'aghy, J. Kroha, L. Szunyogh, and A. Zawadowski,
Theory of the Fano Resonance in the STM Tunneling Density of States due to a Single Kondo Impurity,
Phys. Rev. Lett. \textbf{85}, 2557 (2000).

\bibitem{meri} J. Merino and O. Gunnarsson, 
Role of Surface States in Scanning Tunneling Spectroscopy of (111) Metal Surfaces with Kondo Adsorbates,
Phys. Rev. Lett. \textbf{93}, 156601 (2004).

\bibitem{mirages} A. Aligia, and A. Lobos,  
Mirages and many-body effects in quantum corrals,
J. Phys.: Condens. Matter \textbf{17}, S1095 (2005).

\bibitem{lin} C.-Y. Lin, A. H. Castro Neto, and B. A. Jones, 
First-Principles Calculation of the Single Impurity Surface Kondo Resonance,
Phys. Rev. Lett. 97, 156102 (2006).

\bibitem{trimer} A. A. Aligia,
Effective Kondo model for a trimer on a metallic surface,
Phys. Rev. Lett. \textbf{96}, 096804 (2006).

\bibitem{serge} S. Florens, A, Freyn, N. Roch, W. Wernsdorfer, F. Balestro, P. Roura-Bas and A. A. Aligia, 
Universal transport signatures in two-electron molecular quantum dots:
gate-tunable Hund's rule, underscreened Kondo effect and quantum phase
transitions,
J. Phys. Condens. Matter \textbf{23}, 243202 (2011).

\bibitem{frank} S. Frank and D. Jacob, Phys. Rev. B \textbf{92}, 235127 (2015).

\bibitem{morr} D. K. Morr,
Theory of scanning tunneling spectroscopy: From Kondo impurities to heavy fermion materials,
Rep. Prog. Phys. \textbf{80}, (2017).

\bibitem{interf} P. Roura-Bas, F. G\"uller, L. Tosi and A. A. Aligia, 
Destructive quantum interference in transport through molecules with
electron–electron and electron-vibration interactions,
J. Phys.: Condens. Matter \textbf{31}, 465602 (2019).

\bibitem{lim} J. Fern\'andez, P. Roura-Bas, and A. A. Aligia, 
Theory of Differential Conductance of Co on Cu(111) Including Co s and d Orbitals, and Surface and Bulk Cu States,
Phys. Rev. Lett. \textbf{126}, 046801 (2021).

\bibitem{blesio18} G. G. Blesio, L. O. Manuel, P. Roura-Bas, and A. A. Aligia, 
Topological quantum phase transition between Fermi liquid phases in an Anderson impurity model, 
Phys. Rev.  B {\bf 98}, 195435 (2018). 

\bibitem{blesio19} G. G. Blesio, L. O. Manuel, P. Roura-Bas, and A. A. Aligia, 
Fully compensated Kondo effect for a two-channel spin $S = 1$ impurity, 
Phys. Rev.  B {\bf 100}, 075434 (2019). 

\bibitem{leo} L. De Leo and M. Fabrizio, Spectral properties of a
two-orbital Anderson impurity model across a non-Fermi-liquid fixed point,
Phys. Rev. B 69, 245114 (2004),

\bibitem{curtin18} O. J. Curtin, Y. Nishikawa, A. C. Hewson, and D. J. G.
Crow, Fermi liquids and the Luttinger theorem, J. Phys. Commun. \textbf{2},
031001 (2018).

\bibitem{nishi} Y. Nishikawa, O. J. Curtin, A. C. Hewson, and D. J. G. Crow,
Magnetic field induced quantum criticality and the Luttinger sum rule, Phys.
Rev. B \textbf{98}, 104419 (2018).


\bibitem{nl} R. \v{Z}itko, G. G. Blesio, L. O. Manuel and A. A. Aligia,
Iron phthalocyanine on Au(111) is a ``non-Landau'' Fermi liquid,
Nature Commun. \textbf{12}, 6027 (2021).

\bibitem{sb} A. A. Aligia, 
Low-energy physics for an iron phthalocyanine molecule on Au(111),
Phys. Rev. B \textbf{105}, 205114 (2022).

\bibitem{mohr} M. Mohr, M. Gruber, A. Weismann, D. Jacob, P. Abufager, N. Lorente, and R. Berndt, Spin dependent transmission of 
nickelocene-Cu contacts probed with shot noise, 
Phys. Rev. B \textbf{101}, 075414 (2020).

\bibitem{nc} G. G. Blesio, R. \v{Z}itko, L. O. Manuel, and A. A. Aligia,
Topological quantum phase transition of nickelocene on Cu(100),
SciPost Phys. \textbf{14}, 042 (2023)

\bibitem{hein} A. J. Heinrich, J. A. Gupta, C. P. Lutz, and D. M. Eigler,
Single-Atom Spin-Flip Spectroscopy,
Science \textbf{306}, 466 (2004).

\bibitem{hir} C. F. Hirjibehedin, C.-Y. Lin, A. F. Otte, M. Ternes, C. P. Lutz, B. A. Jones, and A. J. Heinrich, 
Large Magnetic Anisotropy of a Single Atomic Spin Embedded in a Surface Molecular Network,
Science \textbf{317}, 1199 (2007).

\bibitem{loth} S. Loth, M. Etzkorn, C. P. Lutz, D. M. Eigler, and A. J. Heinrich, 
Measurement of fast electron spin relaxation times with atomic resolution,
Science \textbf{329}, 1628 (2010).

\bibitem{paul}  W. Paul, K. Yang, S. Baumann, N. Romming, T. Choi, C. P. Lutz, and A. J. Heinrich, 
A scanning tunneling microscope capable of electron spin resonance and pump-probe spectroscopy at mK temperature and in vector magnetic field,
Nat. Phys. \textbf{13}, 403 (2017).

\bibitem{tri} S. Trishin, C. Lotze, N. Bogdanoff, 
F. von Oppen, and K. J. Franke,
Moir\'e Tuning of Spin Excitations: Individual Fe Atoms on 
MoS$_2$/Au(111),
Phys. Rev. Lett. \textbf{127}, 236801 (2021).

\bibitem{ternes} M. Ternes, 
Spin excitations and correlations in scanning
tunneling spectroscopy,
New J. Phys. \textbf{17}, 063016 (2015).

\bibitem{tri2} S. Trishin, C. Lotze, F. Lohss, G. Franceschi, L. I. Glazman, F. von Oppen, K. J. Franke,
Tuning a two-impurity Kondo system by a moir\'e superstructure,
arXiv:2301.01517.

\bibitem{wang} Y. Wang, B. Wang, R. Huang, B. Gao, F. Kong, and Q.
Zhang,
First-principles study of transition-metal atoms adsorption on
MoS$_2$ monolayer,
Physica (Amsterdam) \textbf{63E}, 276 (2014).

\bibitem{chen} X. Chen, L. Zhong, X. Li, and J. Qi,
Valley splitting in the transition-metal dichalcogenide monolayer via
atom adsorption,
Nanoscale \textbf{9}, 2188 (2017).

\bibitem{tkro} M. Moro-Lagares, J. Fern\'andez, P. Roura-Bas, M. R. Ibarra, A. A. Aligia, and D. Serrate, 
Quantifying the leading role of the surface state in the Kondo effect of Co/Ag(111)
Phys. Rev. B \textbf{97}, 235442 (2018).

\bibitem{zitko09} R. \v{Z}itko and T. Pruschke,
Energy resolution and discretization artifacts in the numerical renormalization group, 
Phys. Rev. B \textbf{79}, 085106 (2009).

\bibitem{nrglj} NRG Ljubljana, \url{https://github.com/rokzitko/nrgljubljana} and \url{http://nrgljubljana.ijs.si/}.

\bibitem{bulla} R. Bulla, T. Costi and T. Pruschke, 
The numerical renormalization group method for quantum impurity systems, 
Rev. Mod. Phys. \textbf{80}, 395 (2008).

\bibitem{joaq} J. Fern\'andez, P. Roura-Bas, A. Camjayi, and A. A. Aligia,
Two-stage three-channel Kondo physics for an FePc molecule on the Au(111) surface,
J. Phys.: Condens. Matter \textbf{30}, 374003 (2018); 
Corrigendum J. Phys. Condens. Matter \textbf{31}, 029501 (2018).



\end{thebibliography}
\end{document}